# Detailed instantaneous ionization rate of $H_2^+$ in intense laser field

**Mohsen Vafaee [1], Hassan Sabzyan [1], Zahra Vafaee [2] Ali Katanforoush [3]**

[1] Department of Chemistry, University of Isfahan, Isfahan 81746-73441, I. R. Iran

[2] Department of Mathematics, University of Isfahan, Isfahan 81746-73441, I. R. Iran

[3] School of Mathematics, Institute for Studies in Theoretical Physics and Mathematics (IPM), Tehran 19395-5746, I. R. Iran

E-mail: 1) MohsenVafaee@sci.ui.ac.ir, 2) sabzyan@sci.ui.ac.ir

**Abstract**

Component of the instantaneous ionization rate (IIR) is introduced and calculated for $H_2^+$ in a linearly polarized laser field with $1.0 \times 10^{14}$ W/cm$^2$ intensity and $\lambda = 1064$ nm wavelength by direct solution of the fixed-nuclei full dimensional time-dependent Schrödinger equation. The component ionization rates, calculated for different values of inter-nuclear distance, are compared with those calculated via the virtual detector method (VD). Details of the time dependent behavior of the outgoing and incoming electron wavepakets of the $H_2^+$ system in intense laser field at attosecond time scale are studied based on the calculated component ionization rates. It has been shown clearly that the positive signals of the IIR (outgoing electron wavepacket signals) are strong and sharp but the negative signals of the IIR (returning electron wavepacket signals) are smooth and weak. It is also shown that for H-H distance R<5.6, when the laser pulse is turned on with a ramp, the R-dependent ionization rates move towards those of the upper time dependent Floquet quasi-energy state (QES).

**Introduction**

In intense laser field, atoms and molecules are ionized in the attosecond scale [1]. In addition to the electronic dynamics, molecules have nuclear dynamics, vibration and rotation, in femtosecond and picosecond time scales. Extensive studies carried out experimentally and theoretically on the $H_2$ and $H_2^+$ systems have revealed comprehensive new phenomena [2-4]. Studies of the dynamics of these two systems exposed to intense laser field are very complicated because involved simultaneously two processes, ionization and dissociation.

An interesting and complicated effect observed experimentally and theoretically is the enhancement of the ionization rate of $H_2^+$ as a function of H-H inter-nuclear separation which results in maxima at some critical points. Indirect [5, 6] and direct [7] evidences have been reported for a similar behavior in more complex molecules. In order to characterize and interpret the observed enhanced ionization rates [8-13], recent theoretical studies on the $H_2^+$–laser interaction have been concentrated on the calculation of the ionization rate as a function of inter-nuclear distance R [14-28]. The enhanced ionization has a very important role in the interpretation of the intra-molecular dynamics and fragmentation in intense laser field [24].

Recently, calculations of the ionization rates have had very important improvements. For the Ti:sapphire laser at intensities starting from just above the Coulomb explosion (CE) threshold, it is shown that the R-dependent ionization peaks moves towards small inter-nuclear distances, and their structures become simpler and smoother with the increase in the intensity of the laser pulse (i.e. with the decrease in the Keldysh parameter) [22]. The virtual detector method made it possible to define and distinguish ionization rates and outgoing norms from different boundaries which can be used to study details of the dynamics of the spatial evolution of the electron wavefunction [18]. Moreover, based on the instantaneous ionization rates (IIR), details of the time-dependent behavior of the system following the variations of the laser field can be extracted and used in the interpretation of the enhanced ionization rates [28]. This approach allows us to determine not only the instantaneous intensity but also the instantaneous direction of the electronic current. This approach provides direct evidence for the existence of the effect of charge-resonance-enhanced multiphoton resonances of the quasienergy states (QES) with excited electronic states at some particular internuclear distances [19,28].

In this article, we present more details about the structure of the enhanced ionization rate of $H_2^+$ as a function of inter-nuclear separation by calculating and analyzing overall and component IIR and the time-averaged overall ionization rates for different values of the rising time of the intense laser pulse. In the rest of this article, the numerical method used for the modeling of the $H_2^+$ system is introduced briefly and then the results obtained for the $\lambda = 1064$ nm and $1.0 \times 10^{14}$ W/cm$^2$ laser light are presented. Next, the components of the IIR are introduced and their results are presented and discussed.

**Numerical solution of the TDSE**

Time dependent Schrödinger equation, TDSE, in the cylindrical polar coordinates for the $H_2^+$ molecular ion exposed to the laser field of $E(t) = E_0 f(t) \cos(\omega t)$ applied along the inter-nuclear axis ($z$ axis) in atomic units reads as [16, 18]

$$i\frac{\partial \psi(z,\rho,t)}{\partial t} = H(z,\rho,t)\psi(z,\rho,t), \tag{1}$$

in which

$$H(z,\rho,t) = -\frac{2m_p + m_e}{4m_p m_e}\left[\frac{\partial^2}{\partial \rho^2} + \frac{1}{\rho}\frac{\partial}{\partial \rho} + \frac{\partial^2}{\partial z^2}\right] + V_C(z,\rho,t) \tag{2}$$

$$V_C(z,\rho,t) = -\sum_{\pm}\left((z \pm R/2)^2 + \rho^2\right)^{-\frac{1}{2}} + \left(\frac{2m_p + 2m_e}{2m_p + m_e}\right)zE_0 f(t)\cos(\omega t) \qquad (3)$$

The laser pulse envelope $f(t)$ is set as

$$f(t) = \begin{cases} \frac{1}{2}\left[1 - \cos\left(\frac{\pi t}{\tau_1}\right)\right] & \text{for } 0 \leq t \leq \tau_1 \\ 1 & \text{for } \tau_1 \leq t \leq \tau_1 + \tau_2 \\ \frac{1}{2}\left[1 - \cos\left(\frac{\pi(t - \tau_2 - 2\tau_1)}{\tau_1}\right)\right] & \text{for } \tau_1 + \tau_2 \leq t \leq 2\tau_1 + \tau_2 \\ 0 & \text{for } 2\tau_1 + \tau_2 \leq t \end{cases} \qquad (4)$$

with $\tau_1$ and $\tau_2$ being the rising time and duration of the laser pulse at its full-scale amplitude, respectively. The differential operators are discretized by the eleven-point difference formulae which have tenth-order accuracies [28]. More details of our calculations are described in our previous reports [18, 28].

**Time-Dependent Ionization Rates; Results and Discussion**

The ionization rate, $\Gamma$, is obtained by calculating time-dependent norm, $N(t)$, of the wavefunction via

$$\Gamma(t) = \frac{-d\ln\left(N(t)/\|\Psi(0)\|^2\right)}{dt} \qquad (5)$$

where $N(t) = \|\Psi(t)\|^2$. Simulation of the time-dependent behavior of the $H_2^+$ system at some fixed inter-nuclear separations in a box of (640,170) size for the $(z,\rho)$ coordinates carried out by direct solution of the TDSE in the presence of the linearly polarized laser field with $I=1.0\times10^{14}$ W/cm$^2$ and $\lambda = 1064$ nm has been reported in our previous report [28]. In this simulation, parameters of the pulse envelope are set to $\tau_1 = 5$ and $\tau_2 = 15$. The overall ionization rate of H$_2^+$ at each inter-nuclear distance is calculated by averaging the calculated instantaneous ionization rates over the period $\tau = 10$ to $\tau = 20$ cycles of the laser field using Eq. (5). Figure 1 shows the R-dependent ionization rates calculated for the H$_2^+$ system (●). In addition to the results reported in our previous paper [28], this figure contains a new set of points at large values of R, above 14.0. The isolated single point on the right vertical axis corresponding to the ionization rate of an isolated H atom in the same laser field is given for comparison. Figure 1 shows that at large internuclear distances, the ionization rates of H$_2^+$ approache those of the isolated H atom. As discussed in [28], the calculated ionization rates, especially above R>9.5, are in good agreement with those reported by others [16,17,19]. It can also be seen from Figure 1 that the R-Γ curve below R=6.5 has a fine structure with both narrow and weak peaks.

We have shown already [18] that in a cylindrical coordinate, it is possible to decompose the outgoing norm of the system as $N_{VD}^O(t) = N_z^O(t) + N_\rho^O(t)$, in which $N_z^O(t)$ and $N_\rho^O(t)$ are respectively the instantaneous outgoing norms from the $z$ and $\rho$ boundaries, so that the overall ionization rate $\Gamma_{VD}$ and its $z$ and $\rho$ component, $\Gamma_z$ and $\Gamma_\rho$, calculated via virtual detector (VD) method, are connected via [18]

$$\Gamma_{VD}(t) \approx \Gamma_z(t) + \Gamma_\rho(t). \tag{6}$$

This means that the total VD ionization rate $\Gamma_{VD}$ is approximately equal to the sum of the component ionization rates for the $z$ and $\rho$ boundaries $\Gamma_z$ and $\Gamma_\rho$, both calculated via the VD method. Figure 2 shows the total VD ionization rate, $\Gamma_{VD}(\bullet)$, and its components, $\Gamma_z$ ($\Delta$) and $\Gamma_\rho$ ($\square$) as functions of internuclear separation R calculated for $H_2^+$ in intense laser field of I=$1.0\times 10^{14}$ W/cm$^2$ and $\lambda$ = 1064 nm in a simulation box of (640,170).

Comparison of Figure 2 and Figure 4a of Ref. [18] shows the wavelength effects of the intense laser field on the total VD ionization rate $\Gamma_{VD}$ and its components ionization rates $\Gamma_z$ and $\Gamma_\rho$. Variations of the component ionization rates are decreased with the increase in the wavelength from 790 nm to 1064 nm.

We can see from Figure 2 that the first peak of the ionization rate at R~4.8 is mainly due to the ionization rate of the z-coordinate $\Gamma_z$, while the second and the third peak, at 6.0 and 9.4, have contributions from both component ionization rates $\Gamma_z$ and $\Gamma_\rho$; comparative contributions of the component ionization rates for the two peaks are respectively $\Gamma_z > \Gamma_\rho$ and $\Gamma_\rho > \Gamma_z$. In this simulation box, for internuclear distances above R=13.0, the two component ionization rates approach one another. As shown in our recent study [18], the relative magnitudes of the component ionization rates $\Gamma_z$ and $\Gamma_\rho$ depend also on the relative sizes of the simulation box along the $\rho$ and $z$ coordinates.

The instantaneous energy of the $H_2^+$ system in intense laser field can be decomposed and calculated as $E(t) = E_{Re} + iE_{Im}$, where $E_{Re}$ and $E_{Im}$ are the real and imaginary parts, respectively. The quantity $\Gamma(t)$ that is defined as:

$$\Gamma(t) = \frac{-2E_{Im}(t)}{N(t)} \qquad (7)$$

is the instantaneous ionization rates (IIR) of the system [28]. The imaginary energy can be decomposed as $E_{Im} = T_{Im,z} + T_{Im,\rho}$, because the potential part of the imaginary energy, $V_{Im}$, is zero and thus has no contribution in $E_{Im}$. Therefore, Eq. (7) can be rewritten as

$$\Gamma(t) = \frac{-2T_{Im,z}(t)}{N(t)} + \frac{-2T_{Im,\rho}(t)}{N(t)} = \Gamma_z(t) + \Gamma_\rho(t) \qquad (8)$$

This equation shows that the overall IIR, $\Gamma(t)$, can be decomposed into its two $z$ and $\rho$ IIR components, $\Gamma_z(t)$ and $\Gamma_\rho(t)$, in the same way as the total VD ionization rate [18].

The IIR and its components for $H_2^+$ at some fixed inter-nuclear separations have been calculated by direct solution of the TDSE in the presence of the linearly polarized laser field with I=1.0×10$^{14}$ W/cm$^2$ intensity and $\lambda$ = 1064 nm wavelength with the envelope parameters $\tau_1 = 5$ and $\tau_2 = 15$. The overall ionization rate of $H_2^+$ at each inter-nuclear distance is calculated by averaging the calculated IIR over the period $\tau = 10$ to $\tau = 20$. Figure 3 shows the R-dependent overall ionization rate $\Gamma(\bullet)$ and its components, $\Gamma_z(\Delta)$ and $\Gamma_\rho(\square)$, calculated in the present study for the $H_2^+$ system. Structures of these R-dependent curves are similar to those of the corresponding curves in Figure 2. While additivity of the VD ionization rate components $\Gamma_z$ and $\Gamma_\rho$, and thus the equality of Eq. (6), is an approximation, the relation between the instantaneous ionization rate $\Gamma(t)$ and its two component $\Gamma_z(t)$ and $\Gamma_\rho(t)$ in Eq. (8) is exact.

On the basis of Eqs. (7) and (8), for $\Gamma(t)$ and its components $\Gamma_z(t)$ and $\Gamma_\rho(t)$, positive (negative) values of instantaneous imaginary energy result in negative (positive) value for instantaneous ionization rates, and therefore, the positive (negative) values of the instantaneous imaginary energy correspond to the incoming (outgoing) of the electron to (from) the system [28].

The distinct advantage of the evaluation of the IIR via Eqs. (7) and (8) is that it makes it possible to follow the time-dependent ionization processes of a system and to study details of the ionization mechanism [28]. To show this ability more clearly, the time-dependent ionization rates of $H_2^+$ in the linearly polarized field of I= $1.0\times10^{14}$ W/cm$^2$ intensity and $\lambda = 1064$ nm wavelength for a number of selected inter-nuclear separations have been calculated and presented in Figure 4. The R values in Figure 4 correspond to the peaks of the R-dependent ionization rates presented in Figure 1, i.e. 4.8, 6.0 and 9.6. These Figures demonstrate the behavior of the IIR and its components over the two cycles (from $\tau = 14$ to $\tau = 16$) of the laser pulse ( 1 cycle ~ 3.55 fs).

Analysis of Figure 4 reveals that the overall pattern and structure of the IIR $\Gamma(t)$ is dominantly determined by the $z$ component IIR, $\Gamma_z(t)$. Figure 4 shows also that during a cycle, the $\Gamma_z(t)$ curves strongly fluctuate besides having a baseline consisting of an ordered oscillation with the same frequency as that of the laser field. Therefore, the sharp peaks of $\Gamma(t)$ can primarily be attributed to the variation of $\Gamma_z(t)$. In contrast to the variation of $\Gamma_z(t)$, variation of $\Gamma_\rho(t)$ has a considerably smaller amplitude. Comparison between the two component ionization curves shows that variation of the outflow of electron from the $\rho$ boundary is relatively much smoother than that from the $z$ boundary. Each base line peak of the $\Gamma_z(t)$ IIR corresponds to one cycle of the laser pulse. The $\Gamma_\rho(t)$ IIR curves does not show any significant oscillating structure, but have distinct smooth peaks for every half cycle of the laser pulse. A general character of the baseline peaks of the $\Gamma_\rho(t)$ is that (unlike those of the $\Gamma_z(t)$) they do not fall to zero suddenly. Furthermore, the baseline peaks of the $\Gamma_\rho(t)$ have significant intensity at R=9.4. Figure 4 shows clearly the details of the comparative trends of the two component IIRs demonstrated in Figure 3 and discussed above.

**Comparison between the results obtained by TDSE and time-dependent Floquet quasienergy approaches**

We recently reported details of the enhanced ionization rates based on the IIR [28]. This approach gives direct evidence of the charge-resonance-enhanced multiphoton resonances of the quasienergy states (QES) with excited electronic states at some particular internuclear distances. Here in this article, we present more details of the contributions of the individual time dependent Floquet QES to the overall ionization rates.

Chu and Chu tried to resolve detailed mechanism of the ionization enhancement phenomenon by calculating ionization rates of the QESs [19]. They applied an ac Floquet calculation to study characteristics and the dynamic behavior of the complex QESs of $H_2^+$ at different values of internuclear distance R. They introduced a complex-scaling generalized pseudo spectral (CSGPS) technique for the determination of the complex QESs for two-center diatomic molecular systems. Chu and Chu's calculations showed that the dominant electron population remains in the $1\sigma_g$ and $1\sigma_u$ states in the laser field of $1\times10^{14}$W/cm$^2$ intensity and $\lambda$ = 1064 nm wavelength. Thus, the two distinctly different groups of QESs, i.e. the lower and upper groups, whose major components are the field-free $1\sigma_g$ and $1\sigma_u$, can be derived. Because of the Floquet symmetry, all of the QESs in the lower (or in the upper) group separated by $2\omega$ (n is an integer) in energy, are in fact physically indistinguishable and contain the same information regarding multiphoton dynamics. Thus, all QESs in the lower (or in the upper) group have identical imaginary energies. They used the dynamical properties of these two sets of QESs to explore the mechanism(s) responsible for the enhanced ionization phenomenon. A detailed analysis of the nature and dynamical behavior of these QESs by Chu and Chu [19] reveals that the ionization enhancement is mainly due to the effect of the charge-resonance-enhanced multiphoton resonances of the $1\sigma_g$ and $1\sigma_u$ states with the excited electronic states at some particular inter-nuclear distances. Chu and Chu results, presented in Figure 5, indicate that the QESs in the lower and upper groups both show a double-peak enhancement feature.

To evaluate individual contributions from the lower and upper sets of QESs to the overall ionization rates, we have reported in Figure 5 the calculated overall ionization rates for different laser pulses with different values of rising time $\tau_1$, and compared them with the isolated ionization rates of these two QESs reported by Chu and Chu [19]. As this figure shows, when the field is turned on with $\tau_1 = 5$ cycles (○), the ionization rates show three peaks, at 4.8, 6.0 and ~9.6. The peaks at 6.0 and 9.6 can be attributed to the peaks of the lower QES. But the first peak at 4.8 is placed between the peaks of the lower and upper QES. When the laser pulse is turned on immediately, i.e. with $\tau_1 = 0$, the calculated ionization rates derived by TDSE (●) correspond to those of the lower QES for $R > 4.8$. Figure 5 shows that the TDSE results for $R < 4.8$ with $\tau_1 = 0$ (●), take a slight distance from the lower QES Floquet curve. This small difference is probably due to the avoided crossing of the upper and lower QES's. On the other hand, when the laser pulse is turned on with a ramp or rising part with increasing $\tau_1$ from zero to $\tau_1 = 5$ (○), $\tau_1 = 10$ (△) and $\tau_1 = 15$ (□) cycles for example, the points of the ionization rates for $R < 5.6$ approach the curve of the upper QES. We can conclude that when the laser pulse is turned on with an initial ramp, both lower and upper QES's contribute to construct the first peak of the ionization rates at R=4.8. For all of the $R < 3.6$ and $R > 5.2$ ranges, the ionization rates are determined by the lower QES and the amount of $\tau_1$ does not have significant effect on the R-dependent ionization rate.

**Instantaneous returning of the electron wavepacket**

As mentioned in the preceding section, positive (negative) value of the instantaneous imaginary energy results in negative (positive) value for IIR, and therefore, the positive (negative) value of the instantaneous imaginary energy corresponds to the incoming (outgoing) of the electron to (from) the system. Designing a simulation to show both negative and positive IIR is a difficult task, because the outgoing and incoming electron wavepakets occur simultaneously and are summed to give the overall IIR. Moreover, the intensity of the positive IIR overcomes that of the negative IIR almost usually. Therefore, it seems that it is experimentally impossible to probe the negative IIR. So far, the negative ionization rate has not been reported or noticed in theoretical and computational works. In this section, we introduce details and structures of both negative and positive IIR and their contributions to the corresponding component and overall IIRs and their averages, and their effect on the R-dependent behavior of the ionization rates.

Calculations of IIR of $H_2^+$ at R=9.6 over the first five cycles of the linearly polarized laser field of I=$1.0\times10^{14}$ W/cm$^2$ and $\lambda = 1064$ nm are repeated but with a pulse shape that is turned on as $E_0 \cos(\omega t)$, i.e. with $\tau_1 = 0$. These calculations of the (instantaneous) ionization rates were carried out for boxes all with (640,170) size for the ($z, \rho$) coordinates, shown Figure 6. Borders of the simulation box are placed such that $-300 \leq z \leq 300$ and $0 \leq \rho \leq 150$ and the absorber regions, the hatched area, are set at $-320 \leq z \leq -300$ and $300 \leq z \leq 320$ for the $z$ and at $150 \leq \rho \leq 170$ for the $\rho$ coordinate [28]. The IIRs are calculated for the simulation boxes shown in Figure 6 based on the electron wavepackets passing the borders of the gray regions. In the present calculations of the IIRs, different simulation boxes are used with different positions of the borders of the gray region which are set respectively at (a) $-300 \leq z \leq 50$, (b) $-300 \leq z \leq 50$, (c) $-300 \leq z \leq 300$ and (d) $-250 \leq z \leq 250$ for the $z$, and $0 \leq \rho \leq 150$ for the $\rho$ coordinate, and finally at (e) $-300 \leq z \leq 300$ for the $z$ and $0 \leq \rho \leq 100$ for the $\rho$ coordinate. The two red bullets near the origin are the H nuclei having a 9.6 distance. Figure 7 demonstrates the calculated IIR's and their components corresponding to the simulation boxes of Figure 6.

Let us now study the IIR for the first simulation box (shown in Figure 6a). When the laser field is turned on, the electron wavepacket starts to outgo from the gray region by passing through the $z$ borders placed at -50 and 50 and through the $\rho$ border placed at 150. Since the $z$ borders are closer to the concentrated part (center) of the unperturbed (t=0) electron wavepacket, we thus expect to detect first the outgoing wavepacket from the $z$ borders much earlier than that from the $\rho$ border. If one of the borders of the gray region is placed closer to (farther from) the nuclei (i.e. the center of the unperturbed wavepacket), intensity of the outgoing wavepacket from that border in the initial stages of the laser pulse becomes also stronger (weaker). Therefore, the outgoing electron wavepacket from the $z$ borders in Figure 6a is expected to be stronger than that from the $\rho$ border. The IIRs demonstrated in Figure 7a are in agreement with this expectation. For each cycle of the laser pulse, there should be two distinct stages in the outgoing wavepacket that passes alternatively from the two $z$ borders, located respectively in the $+z$ and $-z$ directions. This is while the outgoing wavepacket from the $\rho$ border should be smooth, however still periodic (due to the periodic behavior of the pulse), with weak variations.

Figure 7a shows one baseline peak for the overall IIR and its $z$ component for every half cycle of the laser pulse related to the passage of the electron wavepacket through the corresponding $+z$ or $-z$ borders. Intensity of this baseline peak reaches its maximum value over the first few cycles of the laser pulse. The first baseline peak is relatively weak. As can be seen from Figure 7a, the $\rho$ component of the IIR for the Figure 6a setup of the simulation box shows always very weak variations and does not distinctly follow up the laser pulse shape. Outgoing electron from the $z$ borders is strong as compared to that from the $\rho$ border, which is weak. In this case, the total ionization rate can be considered to be mainly due to the ionization through the $z-$ borders.

The $+z$ and $-z$ borders in Figure 6b are placed at +50 and -150, respectively. This situation of the $z-$ borders results in two effects. First, the $-z$ component IIR will be vanished for some initial half cycles of the laser pulse due to the longer time needed for the electron wavepackets to reach the $-z$ border than to reach the $+z$ border. Second, the intensity of the $z$ component IIR corresponding to the outflow of electron from the $+z$ border at +50 is very stronger than that of the $-z$ border at -150 which is due to the fact that the $+z$ border is closer to the center of the initial wavepacket. These two effects result in the baseline peaks of the $z$ component IIR with alternative intensities; the alternate strong positive peaks corresponding to the half cycles in which the outflow is towards the $+z$ direction and the alternative negative weak peaks corresponding to the half-cycles in which the outflow is towards the $-z$ direction. A comparison between Figures 7a and 7b shows that the outgoing electron from the $+z$ border at +50 overcomes the incoming electron from the $-z$ border at -150 when the electric field is in the $+z$ direction, but the outgoing electron from the $-z$ border at -150 cannot overcome the incoming electron from the $+z$ border at +50 when the electric field is in the $-z$ direction.

Therefore, the negative and positive IIR peaks alternatively appear along each other in Figure 7b. It is only with this design of the simulation box, Figure 6b, that we can clearly observe the negative IIR. Figure 7b exhibits that the positive signals of the IIR (outgoing electron wavepacket signals) are strong and sharp but the negative signals of the IIR (returning electron wavepacket signals) are smooth and weak.

As the borders are moved sufficiently farther away from the nuclei (as in Figure 6c), intensity of the both positive and negative signals of the IIR become weaker at earlier time stages of the laser pulse, and probability of the returning electron wavepacket (the negative IIR) becomes smaller, Figure 7c. Comparison of Figures 7b and 7c shows that the negative IIR peaks are almost vanished for the simulation box of Figure 6c; only positive IIR peaks appear in Figure 7c. In addition, the positive peaks corresponding to the $+z$ border become weaker in Figure 7c as compared to those in Figure 7b. In Figure 7c, when the electric laser field is in the $+z$ $(-z)$ direction, the outgoing electron from the $+z$ $(-z)$ border overcomes the incoming electron from the $-z$ $(+z)$ border. This, therefore, results in the IIR signals which are always positive.

To study the effect of the region boundary on the IIR signals and its components, we carried out two other sets of calculations based on the simulation boxes demonstrated in Figures 6d and 6e which result are presented in Figures 7d and 7e, respectively. Comparison of Figures 7d and 7e with Figure 7c shows that for the present setup of the simulation box, positions of the box boundaries have insignificant effect on the IIR signals. The difference between Figures 7c, 7d and 7e can be interpreted based on the position of the boundaries according to the discussions presented above, i.e. the closer the borders to the center of the initial wavepacket, the stronger the corresponding component IIR peaks. It is thus shown here that calculation of the IIR and its component is easily possible for any local area.

Figure 7 shows that the structure of the IIR peaks is determined mainly by its $z$ component, and that the $\rho$ IIR component has smooth structure. Furthermore, relative magnitudes of the IIR components depend strongly on the relative positions of the corresponding borders with respect to the center of the initial wavepacket.

**Conclusion**

We introduced and calculated exact component instantaneous ionization rates (IIR) based on the instantaneous values of imaginary energy. It is also possible to calculate IIR via the calculation of instantaneous imaginary energy or instantaneous norm of the system, but the exact evaluation of the component IIR is possible only with the calculation of component instantaneous imaginary energy. It is shown that the instantaneous behavior of the electron wavepacket in intense laser field can be probed closely using the IIR of the system. Moreover, details of the electron wavepacket evolution in different directions of the working space under the influence of the strong ultrashort laser field can be studied by calculating components of the IIR. Analyzing of the IIR signals and their components, allows probing details of the behavior of the outgoing and incoming electron wavepackets of the system in intense laser field in attosecond time scale. In addition, it is also possible to follow the wavepackets in the atomic unit spacetime [29]. It is shown that the positive signals of the component and overall IIR (outgoing electron wavepacket signals) are strong and sharp but the negative signals of the component and overall IIR (returning electron wavepacket signals) are smooth and weak.

We also showed in this report that the TDSE and ac Floquet calculations together form a powerful tool to derive detailed mechanism of the ionization enhancement phenomenon. The comparison between the results obtained by these two methods leads us to determine the individual contributions of the QESs to the overall ionization rates. It is revealed by the present results of the solution of TDSE that when the pulse of the laser field is turned on slowly (with a ramp), both lower and upper QES contribute to construct the first peak of the ionization rates at R=4.8 For all of the $R<3.6$ and $R>5.2$ ranges, the ionization rates are determined by the lower QES, and the duration of the rising time of the laser pulse, i.e. $\tau_1$, does not have any significant effect on the contribution of the two QES for the $R<3.6$ and $R>5.2$ ranges.

**Acknowledgment**

We would like to thanks the University of Isfahan for financial supports and research facilities. We should also acknowledge Scientific Computing Center of the School of Mathematics, Institute for Studies in Theoretical Physics and Mathematics (IPM) of I. R. Iran for providing their High Performance Computing Cluster.

**Figure Captions:**

**Fig. 1.** The ionization rate, $\Gamma$, of $H_2^+$ in the linearly polarized laser field of I= $1.0 \times 10^{14}$ W/cm$^2$ intensity and $\lambda$ = 1064 nm wavelength, averaged from $\tau = 10$ to $\tau = 20$ cycles, as a function of R calculated in this work (●) compared with the calculated ionization rates reported by Lu and Bandrauk (Δ) [16], Peng et al. (□) [17] and Chu and Chu (×) [19]. The isolated single point on the vertical coordinate at R=17 corresponds to the ionization rate of an isolated H atom [28].

**Fig. 2.** Calculated ionization rates of $H_2^+$ in the linearly polarized laser field of I= $1 \times 10^{14}$ W/cm$^2$ intensity and $\lambda$ = 1064 nm wavelength using virtual detector (VD) method; the total VD ionization rate, $\Gamma_{VD}$ (●), and its component ionization rates, $\Gamma_z$ (Δ) and $\Gamma_\rho$ (□).

**Fig. 3.** R-dependent overall, $\Gamma$ (●), and component, $\Gamma_z$ (Δ) and $\Gamma_\rho$ (□), ionization rates of the $H_2^+$ system averaged from $\tau = 10$ to $\tau = 20$ cycles of the linearly polarized laser field of I= $1 \times 10^{14}$ W/cm$^2$ intensity and $\lambda$ = 1064 nm wavelength with the envelope parameters $\tau_1 = 5$ and $\tau_2 = 15$, calculated in the present study using Eq. (9). Compare this Figure with Figure 2.

**Fig. 4.** The instantaneous ionization rates of $H_2^+$ in the linearly polarized laser field of I= $1.0 \times 10^{14}$ W/cm$^2$ intensity and $\lambda$ = 1064 nm wavelength over the two cycles, from $\tau_1 = 14.0$ to $\tau_2 = 16.0$ of the laser pulse with the envelope parameters $\tau_1 = 5$ and $\tau_2 = 15$.

**Fig. 5.** A comparison between the overall ionization rates for different rising times of the laser pulse, $\tau_1$, obtained in this work; the laser field are turned on with a ramp having a length of $\tau_1 = 0$ (●), $\tau_1 = 5$ (○), $\tau_1 = 10$ (Δ) and $\tau_1 = 15$ (□) cycles, respectively. The individual contributions of the upper and lower quasienergy states (QESs) reported by Chu and Chu [19] are plotted for comparison.

**Fig. 6.** The different sub-boxes (highlighted in gray and bordered in red) designed for probing the evolution and ionization of the wavepacket of the electron in a simulation box of $-300 \leq z \leq 300$ and $0 \leq \rho \leq 150$. The hatched strips show the absorber regions. The two red points near the origin represent the two H nuclei located symmetrically around the origin with a 9.6 distance in this case. These boxes span over (a) $-50 \leq z \leq 50$, (b) $-300 \leq z \leq 50$, (c) $-300 \leq z \leq 300$ and (d) $-250 \leq z \leq 250$ all with $0 \leq \rho \leq 150$, and over (e) $-300 \leq z \leq 300$ with $0 \leq \rho \leq 100$. The instantaneous ionization rates calculated for these sub-boxes are presented in Figure 7.

**Fig. 7.** The instantaneous ionization rates corresponding to the sub-boxes introduced in Figure 6. The simulations are carried out for R=9.6 for the first five cycles of the linearly polarized laser pulse of I= $1.0 \times 10^{14}$ W/cm$^2$ intensity and $\lambda = 1064$ nm wavelength which is turned on immediately as $E_0 \cos(\omega t)$, i.e. with $f(t) = 1$.

**Figure 1**

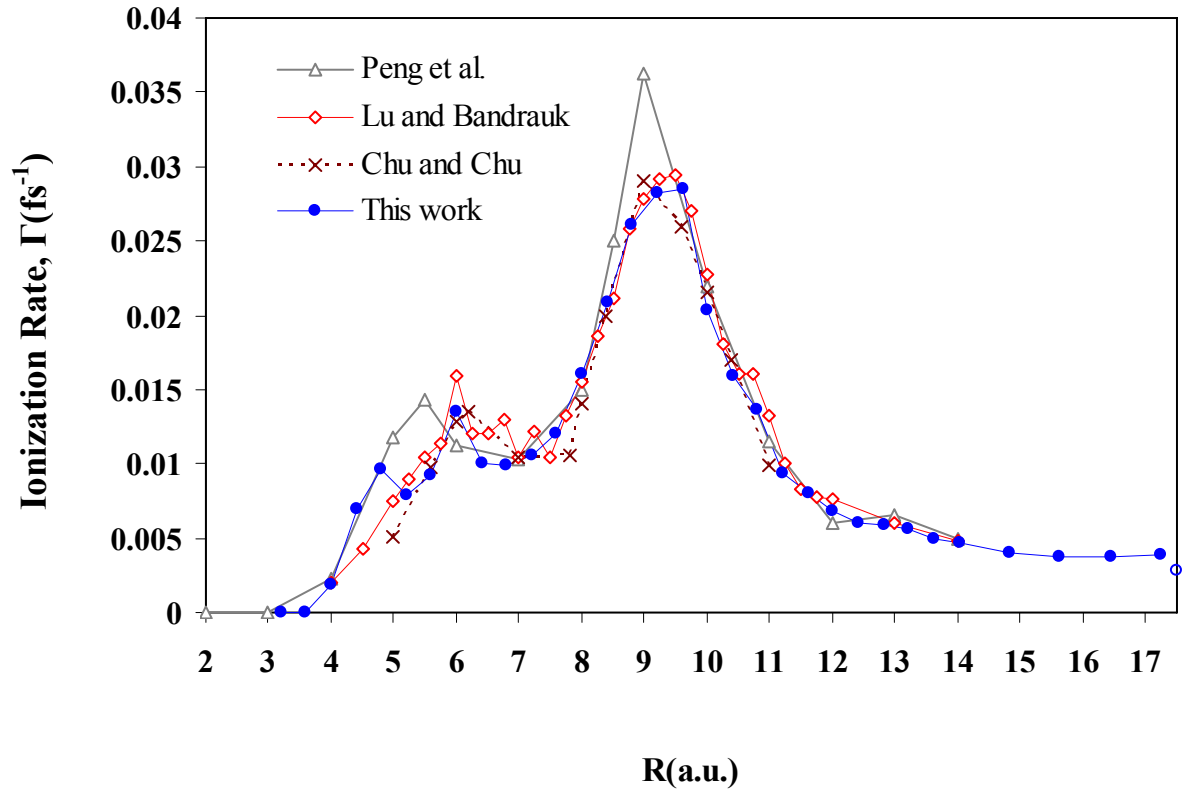

**Figure 2**

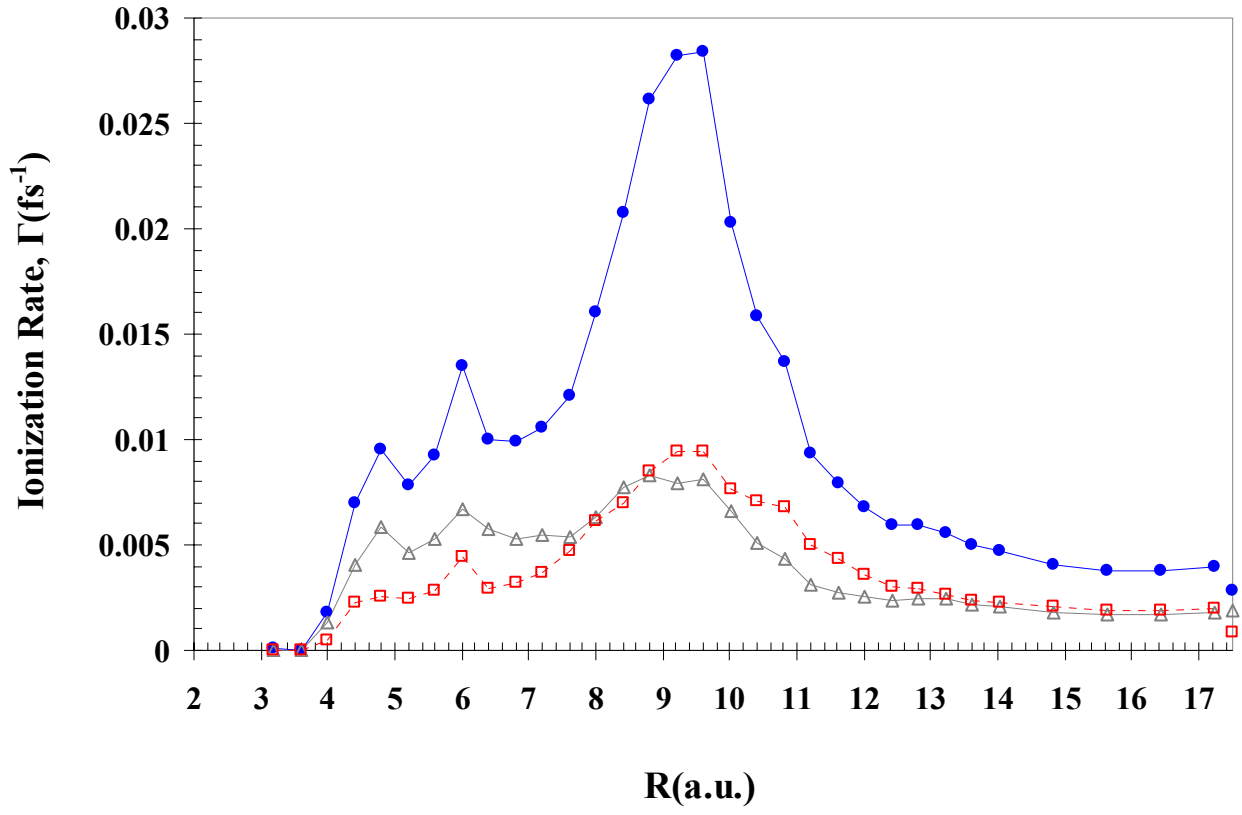

**Figure 3**

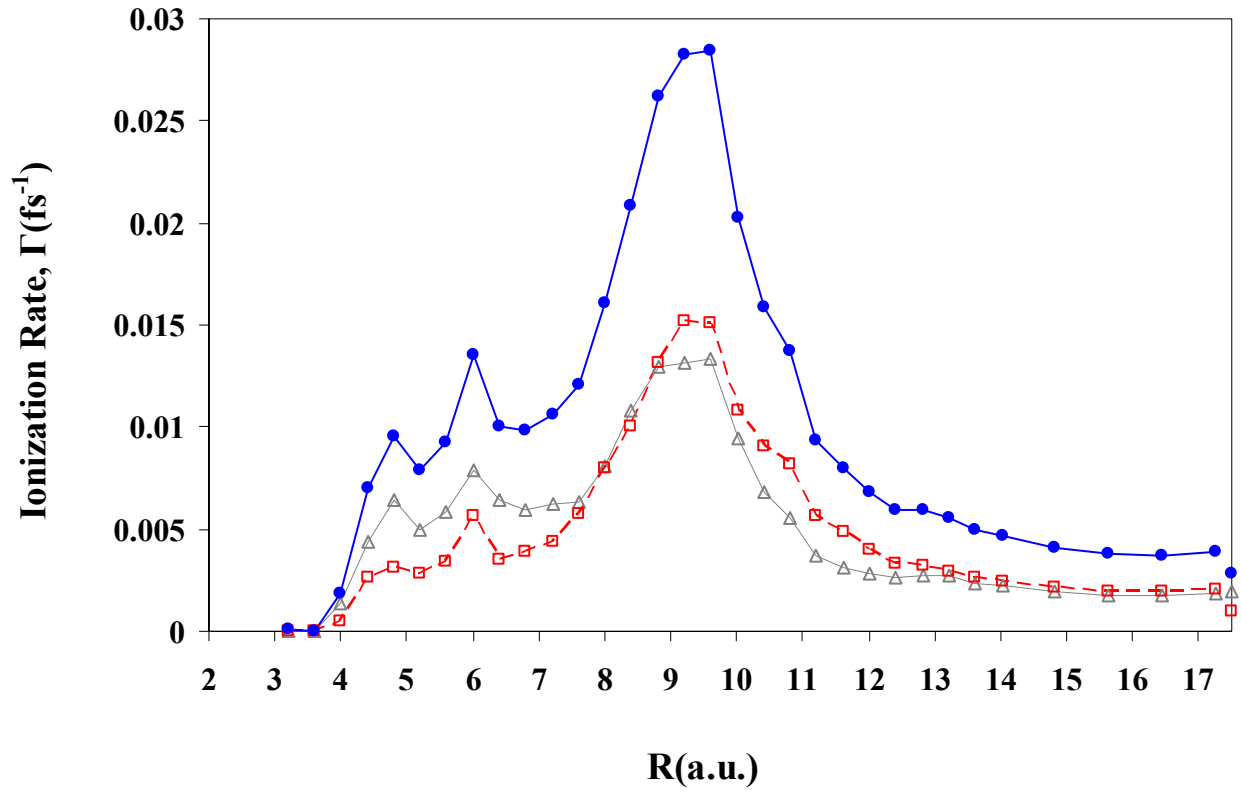

**Figure 4**

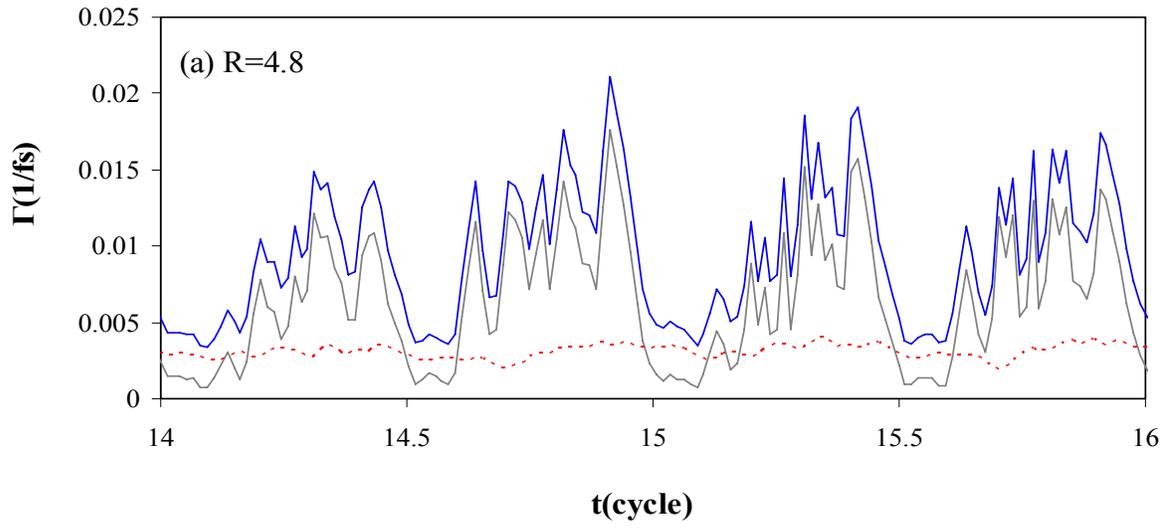

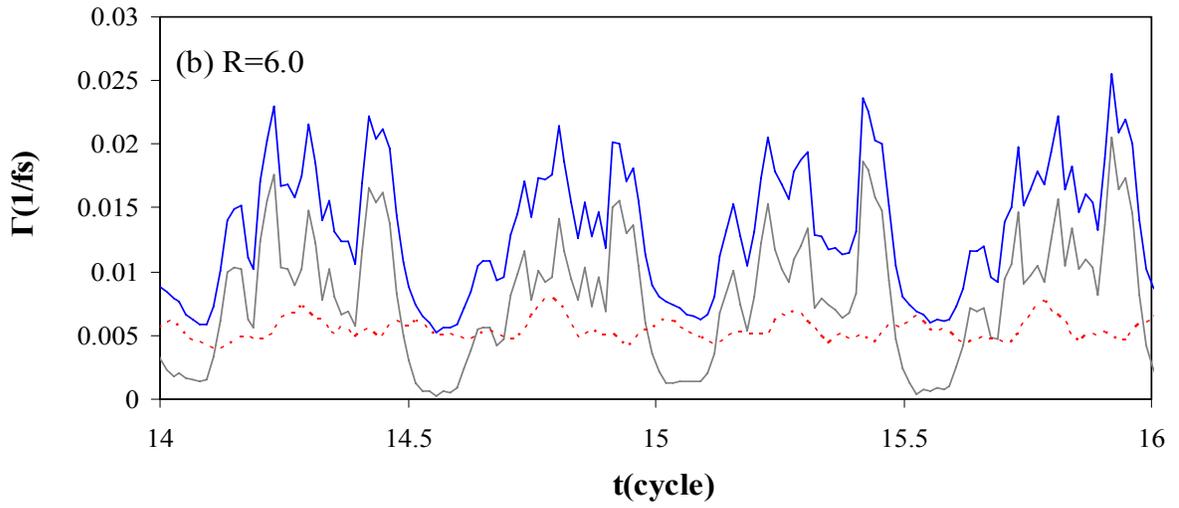

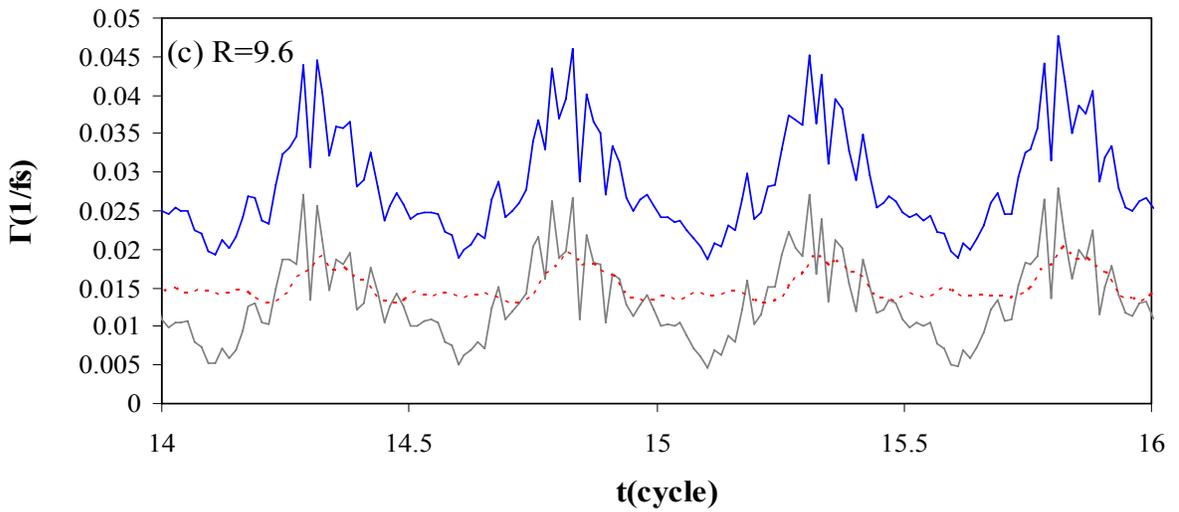

**Figure 5**

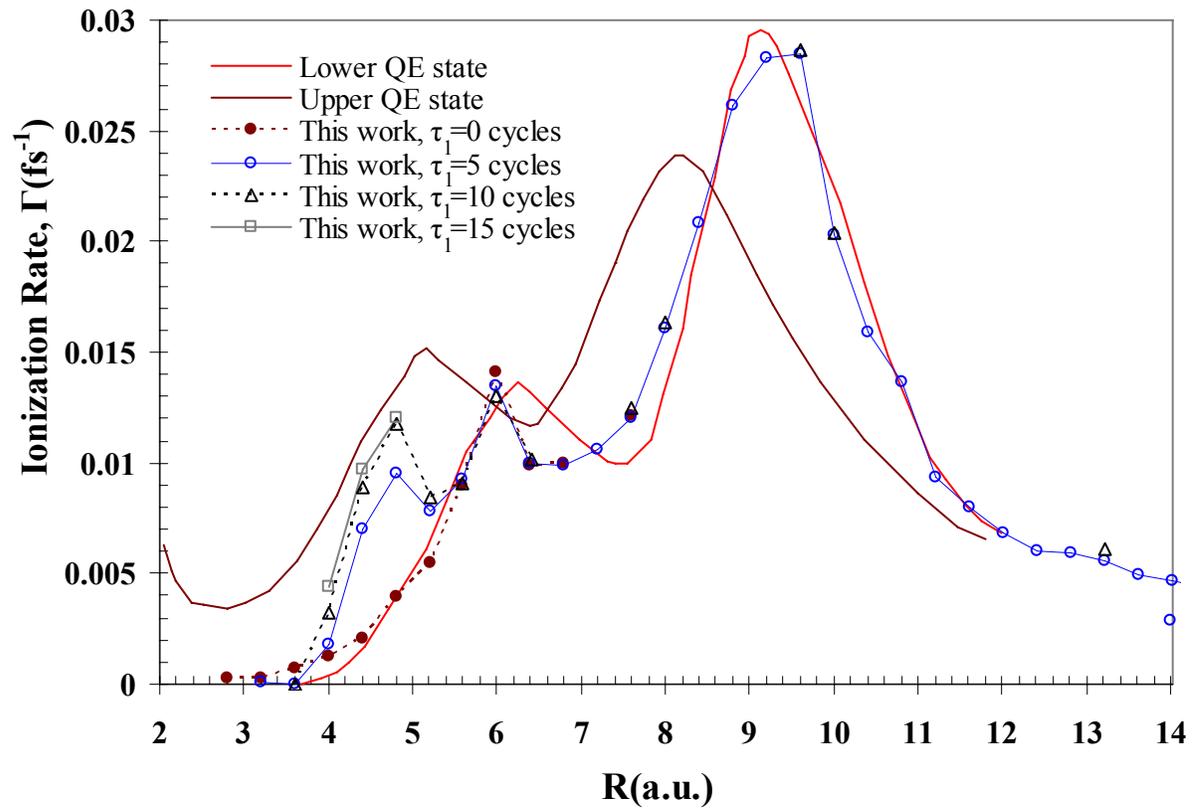

**Figure 6**

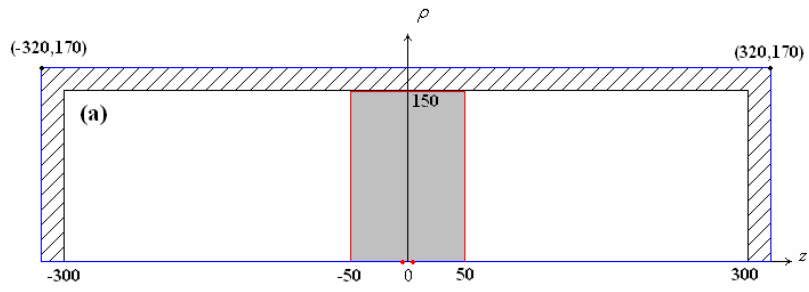

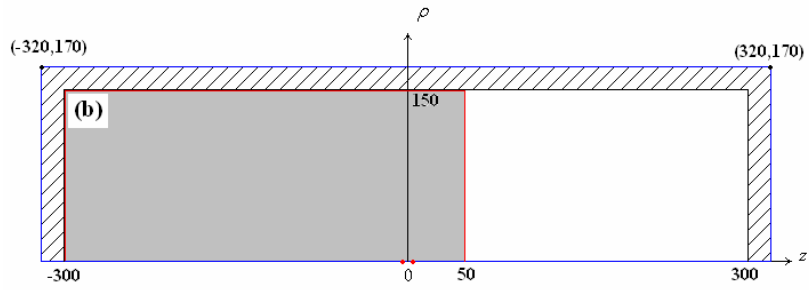

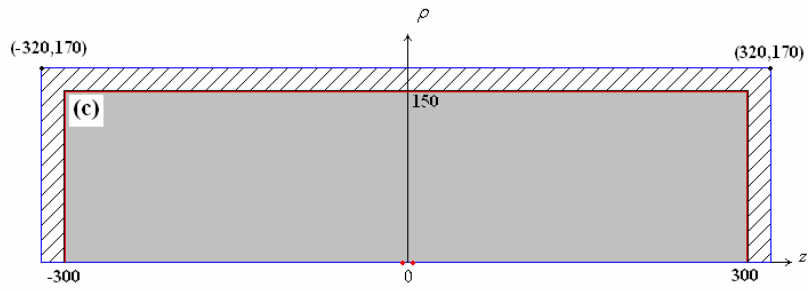

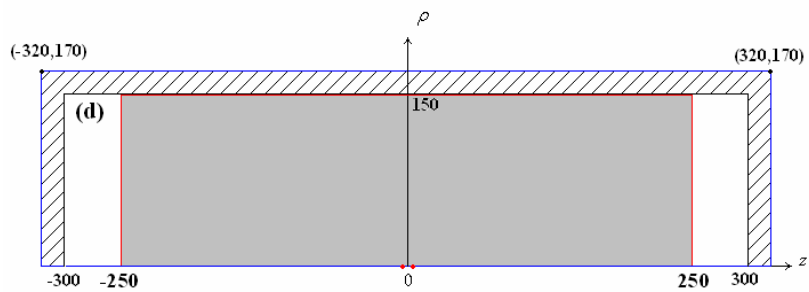

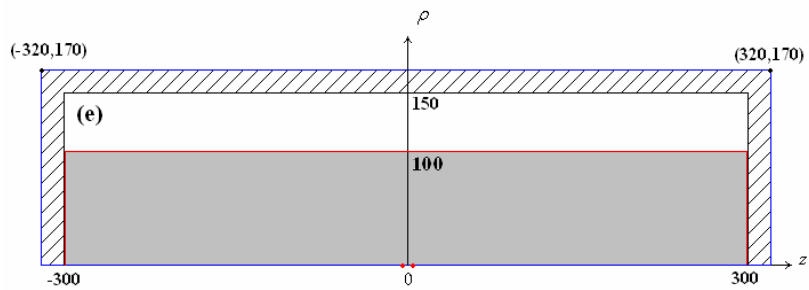

**Figure 7**

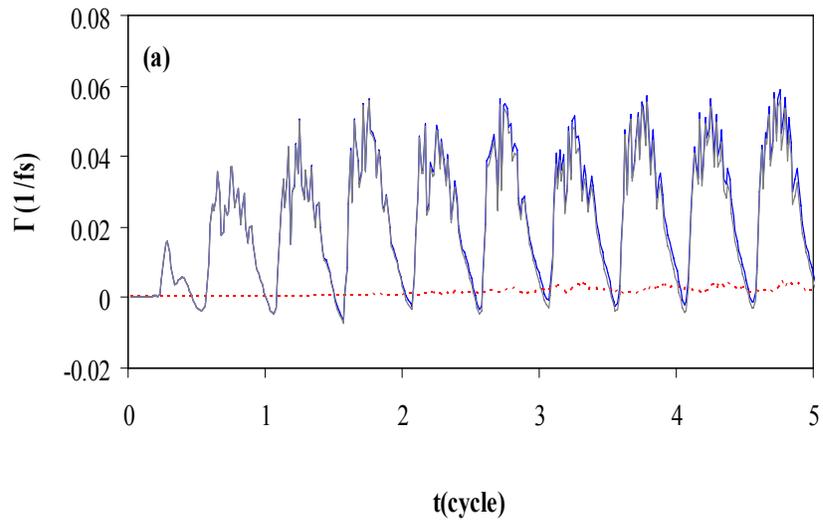

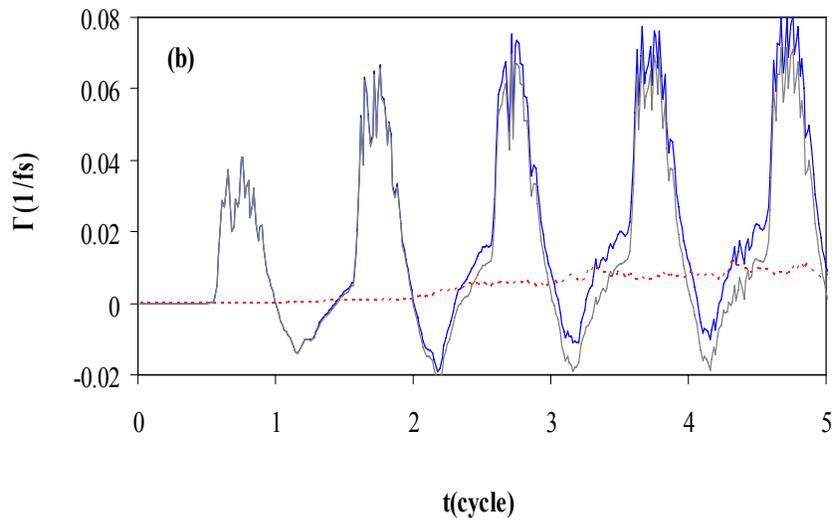

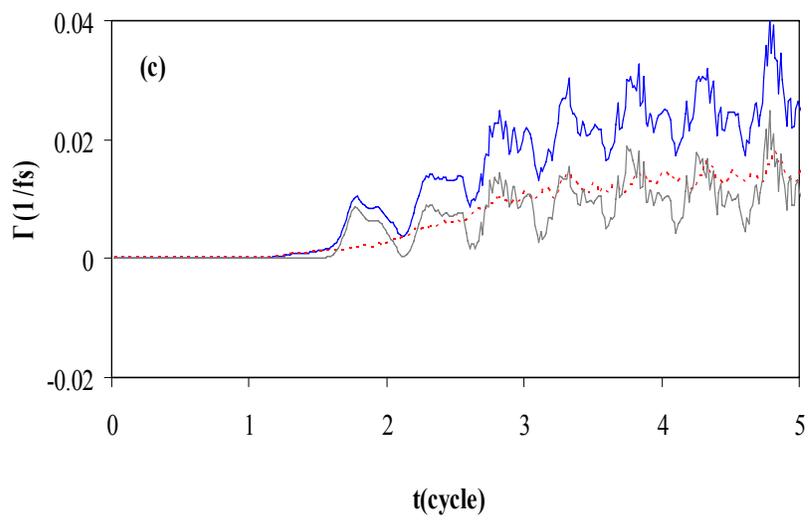

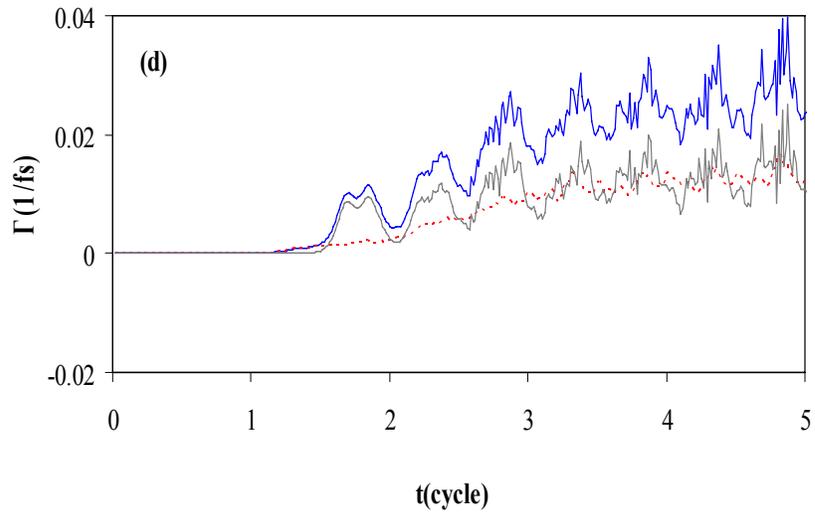

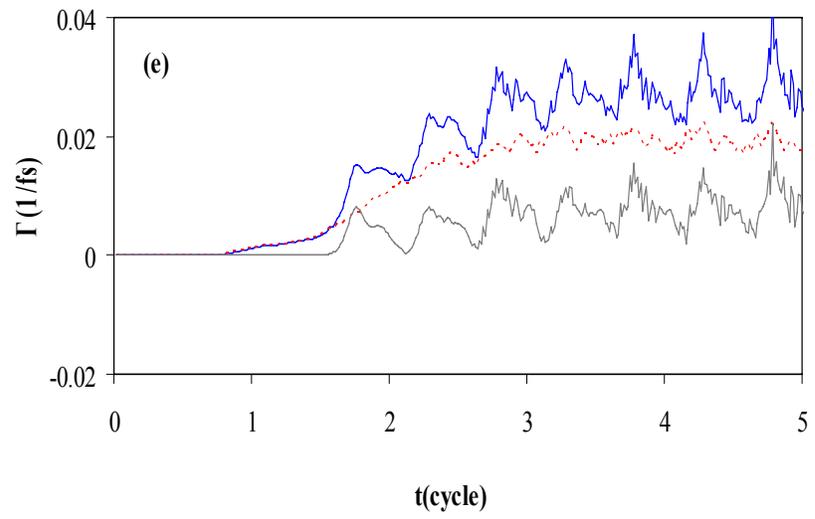